# Reply: Early-onset phenotype of bi-allelic *GRN* mutations

Madam, Sir,

Neuray *et al.* report in this work a series of five new patients from four unrelated families with bi-allelic mutations of *GRN*. This work nicely completes the few existing reports of similar cases, and refers to our recent publication describing six homozygous *GRN* pathogenic variant carriers with divergent phenotypes and ages at onset (Huin *et al.*, 2020). This study provides solid data about clinical features of early-onset bi-allelic *GRN* mutations. All heterozygous pathogenic *GRN* variants reported here were previously associated with frontotemporal dementia (FTD) (Rademakers *et al.*, 2007; Gilberti *et al.*, 2012; Pires *et al.*, 2013; Rovelet-Lecrux *et al.*, 2008; Gijselinck *et al.*, 2008; Clot *et al.*, 2014). The authors also provided follow-up data from previous cases and precise phenotype descriptions and information about progression of the disease. All together, these reports allow better delineation of the clinical spectrum of CLN11 due to pathogenic *GRN* variants.

This present study summarizes the clinical features and evolution of early-onset CLN11. We agree with almost all key-points described here. In particular, the correlation between more severe cognitive deterioration with generalized tonic-clonic seizures (and also pharmacologically refractory epilepsy) is a major point associated with a gene responsible of a pure dementia phenotype elsewhere. Epilepsy is frequent among CLN (Mole *et al.*, 2019) and can correspond to progressive myoclonus epilepsy subtype, characterized by myoclonus, epilepsy with frequent tonic-clonic seizures often pharmaco-resistant, and progressive neurological deterioration (Delgado-Escueta *et al.*, 2001). In CLN11, some cases present myoclonus (Smith *et al.*, 2012; Canafoglia *et al.*, 2014; Almeida *et al.*, 2016; Kamate *et al.*, 2019; Huin *et al.*, 2020) and/or an epileptic syndrome resembling progressive myoclonus epilepsy (Kamate *et al.*, 2019). It would be interesting to know if the five new patients reported by Neuray *et al.* had (or not) myoclonus, considering their similar pharmacologically refractory epilepsy with disabling evolution. This could be important to better characterize epilepsy associated with CLN11, its prognosis and possible treatment.

Neuray *et al.* emphasize the frequency of focal occipital seizures with visual signs in CLN11. Progressive myoclonus epilepsy with focal occipital seizures described in CLN11 patients is similar to that observed in Lafora disease (OMIM #254780) (Turnbull *et al.*, 2016). More broadly, the phenotype of Lafora disease characterized by association of epilepsy, visual loss and cognitive deterioration resembles CLN11. The suspicion of Lafora disease and progressive myoclonus epilepsy with occipital seizures might be a broader indication to measure progranulin plasma level.

As mentioned by the authors, visual symptoms are frequent in patients with bi-allelic *GRN* variants, even if we do not fully agree with the assumption that all the visual symptoms can be considered as red flags for *GRN* testing. Indeed, visual loss leading to progressive blindness is a frequent sign, not specific of CLN11, and even a cardinal feature among most forms of ceroid lipofuscinosis (Mole *et al.,* 2019). We consider more specifically visual hallucinations as red flags in CLN11 patients (Huin *et al.,* 2020). We also note that in the current follow-up of previously published CLN11 patient, one of them reported by Smith *et al.,* developed visual hallucinations (Smith *et al.,* 2012).

Interestingly, two recurrent *GRN* pathogenic variants have already been reported elsewhere at a homozygous state leading to CLN11: the variations c.813_816del, p.Thr22Serfs*10 (Smith *et al.,* 2012; Canafoglia *et al.,* 2014; Neuray *et al.,* 2020) and c.768_769dup, p.Gln257Profs*27 (Faber *et al.,* 2017; Huin *et al.,* 2020; Neuray *et al.,* 2020). Homozygous for p.Thr22Serfs*10 carriers presented first signs at age 22-23 years versus. 15 years, i.e. later than expected. Homozygous variation p.Gln257Profs*27 carriers, reported by us and others, presented with gait impairment or seizures rather than vision loss (Huin *et al.,* 2020; Faber *et al.,* 2017). The comparison of these cases illustrates the phenotypic variability occurring in patients despite a complete loss of progranulin.

In summary, the Letter from Neuray *et al.,* reports valuable findings that lead to better define CLN11 due to bi-allelic *GRN* pathogenic variants. Despite the small sample number that does not allow statistical analysis, the authors underlined the occurrence of cognitive deterioration and epilepsy. Further study of the CLN11 families with functional brain imaging and neuropsychological examinations may be highly informative for the understanding and the clinical characterization of this rare disease.

**Authors:**

Vincent Huin, MD, PhD[1,2], Mathieu Barbier, PhD[1], Alexandra Durr, MD, PhD[1], Isabelle Le Ber, MD, PhD[1,3]

[1] Sorbonne Université, Paris Brain Institute, APHP, INSERM, CNRS, Paris, France

[2] Univ. Lille, Inserm, CHU Lille, U1172 - LilNCog (JPARC) - Lille Neuroscience & Cognition, F-59000 Lille, France

[3] AP-HP, National Reference center "rare and young dementias", IM2A, Pitié-Salpêtrière University Hospital, Paris, France

**ORCID number:**

Vincent Huin = 0000-0001-8201-5406

Mathieu Barbier = 0000-0002-5154-2163

Alexandra Durr = 0000-0002-8921-7104

Isabelle Le Ber = 0000-0002-2508-5181

**Competing interests**

The authors report no competing interests.

**Funding**

This work received funding from the VERUM foundation and 'Investissements d'avenir' ANR-11-INBS-0011 – NeurATRIS: Translational Research Infrastructure for Biotherapies in Neurosciences. This work was funded by the Programme Hospitalier de Recherche Clinique (PHRC) FTLD-exome (to I.L.B., promotion by Assistance Publique – Hôpitaux de Paris) and PHRC Predict-progranulin (to I.L.B., promotion by Assistance Publique – Hôpitaux de Paris).